\documentclass{article}
\pdfoutput=1
\usepackage[utf8]{inputenc}
\usepackage{amsmath}
\usepackage{xcolor}
\usepackage{comment}
\usepackage{multirow}
\usepackage{adjustbox}
\usepackage{graphicx}
\usepackage{indentfirst}
\usepackage{microtype}
\usepackage{float}
\usepackage{newverbs}
\usepackage{appendix}
\usepackage{caption}
\usepackage{subcaption}
\usepackage{lettrine}
\usepackage{verbatim}
\usepackage{authblk}
\usepackage{amsfonts}
\usepackage{url}
\graphicspath{{images/}}

\title{A comparative study of Bot Detection techniques methods with an application related to Covid-19 discourse on Twitter.}

\newsavebox\affbox
\author[1]{Marzia Antenore\thanks{marzia.antenore@uniroma1.it}}
\author[2]{José M. Camacho-Rodríguez \thanks{camachorodriguez.1848773@studenti.uniroma1.it
}}
\author[2]{Emanuele Panizzi\thanks{panizzi@uniroma1.it}}
\affil[1]{Department of Communication and Social Research, Sapienza University of Rome}
\affil[2]{Department of Computer Science, Sapienza University of Rome}
\date{}
\begin{document}

\maketitle

\begin{abstract}
Bot Detection is an essential asset in a period where Online Social Networks(OSN)
is a part of our lives. This task becomes more relevant in crises, as the
Covid-19 pandemic, where there is an incipient risk of proliferation of social bots,
producing a possible source of misinformation. In order to address this issue, it has
been compared different methods to
detect automatically social bots on Twitter using Data Selection. The techniques utilized to elaborate the
bot detection models include the utilization of features as the
tweets’ metadata or the Digital Fingerprint of the Twitter accounts. In addition, it
was analyzed the presence of bots in tweets from different periods of the first months of the Covid-19 pandemic, using the bot detection technique which best fits the scope of the task. Moreover, this work includes also analysis over aspects regarding the discourse of bots and humans, such as sentiment or hashtag utilization.    
\end{abstract}

\section{Introduction } 

According to \cite{gehl2016socialbots}, a bot is a socio-technical entity based on a software program whose aim is to simulate human behavior in Online Social Networks(OSN) such as Facebook, Twitter, or Instagram. Bots are configured to resemblance as humans not just to other human users, but also to the OSN platform \cite{antenore}. Through different methods such as Artificial Intelligence (AI), bots interpret the situation and react accordingly \cite{antenore}.  These entities can cause malicious effects as influencing in changing the online practices of human users and their practices in Social Networks\cite{gehl2016socialbots}, producing
a detrimental impact on politics. There is proof that social bots are crucial in the propagation of fake news and misinformation \cite{abokhodair2015dissecting} 
\cite{forelle2015political}\cite{ferrara2017disinformation} \cite{ratkiewicz2011detecting}.
Moreover, as the bots improve how to simulate the
human behavior, the line between the human user and this socio-technical entity becomes less clear \cite{antenore}, causing concern in the participation of bots in political events because of the negative effect on the quality of democracy \cite{mustafaraj2010obscurity}. This fact has motivated the development of many bot detection techniques during the last few years\cite{alvisi2013sok}, not always being successful in completely solving the problem \cite{antenore}.

This work focuses on Twitter. Some studies estimated that around 15$\%$ of the accounts on Twitter operates automatically or semi-automatically\cite{ferrara2016rise}.  One reason which might have stimulated the rise of the number of bots is the characteristics of Twitter\cite{antenore}. Moreover, it is worth mentioning that a bot in Twitter is regarded as a credible source of information \cite{edwards2014bot}.  In addition to this, bot operated accounts can be more 2.5 times more influential than human-operated accounts \cite{rizoiu2018debatenight}. The two previous facts combined with the capacity of bots to impersonate themselves as humans might produce events that impact politics negatively influencing public opinion, and thus, affecting drastically democratic processes \cite{bessi2016social}. In particular, a significant amount of bots has been used as fake followers of politicians to generate a false impression of popularity \cite{forelle2015political} or utilized by regimes to spread propaganda \cite{howard2016bots}. Other studies show that social bots influenced discourse in social media during Brexit Referendum \cite{bastos2019brexit}, 2017 French presidential election campaign\cite{ferrara2017disinformation}, 2016 US Presidential Election \cite{howard2016bots}, or 2014 Venezuelan protest \cite{forelle2015political}. Another research also displays that bots influenced the public discourse regarding climate change \cite{ross2019social}. 

This research is developed in the context of Covid-19 pandemic, a situation which have concluded in social and economic disruption, apart from the worst economic downturn since the Great Depression \cite{TheGreat18:online}. In addition, work, public events, sports, conferences and education system have been greatly affected by social distancing measures who forced people out of their comfort daily routines and face-to-face interactions. Social Networks such as Twitter have become fundamental to allows people to stay connected and to share information, opinions and reactions around COVID-19. As social interaction moves more and more online, it becomes crucial to study the activity of automated accounts that could alter public debate on central issues such as government policy, public health and individual decision-making in an undesirable fashion. Furthermore, many studies show that bots accounts play a crucial role in the spread of misinformation in Twitter\cite{TheGreat18:online}. As a consequence, spotting the bots is the first step in order to implement measures to protect the quality of democratic processes. 

At the time of this writing, there are already many studies that have analyzed the public discourse on the Covid-19 pandemic on social network sites \cite{chen2020tracking}. Some of them looked at emotional and sentiment dynamics on social media conversation around pandemic related topics\cite{kleinberg2020measuring}\cite{gao2020mental}. Others have focused primarily on bot accounts detection aiming to describe their behavior, in contrast with human activity, and their focal topics of discussion \cite{ferrara2020types}.

In this work, we provide the following contributions: First and foremost, it is made a comparison between supervised bot detection methods from literature, using the metadata of a Twitter account as well as extracting information from the Social Fingerprint of the accounts using compression statistics. Besides, these methods has been implemented using the data selection technique, in which it will be found a subset of training data which provides a consistent model with the best balance for cross validation and cross-domain generalization\cite{pasricha2019detecting}. The methods implemented will be compared with Botometer v3, which was available until September 2020 and it was used in several studies \cite{rauchfleisch2020false}. In addition, it was analysed the presence of bots in tweets from different periods of the first months of the Covid-19 pandemic, using the bot detection technique which best fits the scope of the task. Moreover, this work includes also analysis over other aspects as the distribution of  bots and differences on the discourse between bots and humans based on the sentiment and hashtags.

\textbf{Roadmap.} In \textbf{Chapter 2}, we comment on the literature reviewed to develop this work and summarize its contributions. In \textbf{Chapter 3}, we make a comparison between the approaches presented in \cite{yang2020scalable} and \cite{pasricha2019detecting}, implementing a data selection technique for both of them and using several classification algorithms. Moreover, the bots and human accounts are depicted utilizing some of the features computed for prediction. Eventually, the models implemented are compared with Botometer version 3. In \textbf{Chapter 4}, we analyze the presence of bots in specific periods of the first months of the pandemic. Then, we study differences in the sentiment between bots and humans in the periods studied. In \textbf{Chapter 5}, we discuss some points about the research and draw some conclusions.

\section{Literature review}

Political manipulation for social bots has occurred worldwide, provoking an increasing interest in bot detection in the last decade \cite{cresci2020decade}. Along this time, both supervised and unsupervised techniques have been implemented to overcome this task. Unsupervised methods are more robust than supervised ones because they do not rely on ground truth quality. Research in \cite{jiang2016catching} introduces CATCHSYN, an unsupervised bot detection algorithm based on a Graph mining approach. This technique allows capturing bots through measures of normality and synchronicity, which allows detecting rare and synchronized behaviors. The advantages of this algorithm are scalability and no need for parameters or labeled data. CATCHSYN presents linear-complexity regarding the graph size and only makes use of topology features for the detection.
 The research in \cite{mazza2019rtbust} also presents an unsupervised method. It uses features extracted from the retweeting patterns of the accounts. These features are used with a clustering algorithm to distinguish between bots and humans. Besides, they introduce RTT plots, an informative visualization to observe suspicious behaviors in the retweeting patterns of Twitter accounts. These visualizations need less information than others proposed in literature like \cite{jiang2016catching} and \cite{giatsoglou2015retweeting}.

Supervised methods, though they might have generalization issues, are extensively used for bot detection \cite{cresci2020decade}. In \cite{varol2017online}, it is presented a supervised method with more than 1000 features related to user metadata, friends, network, temporal, content, and sentiment. This research concluded in the first version of Botometer, a bot detection service available online. \cite{yang2019arming} presents an update of that version.  This update added new features to the model and included new training datasets containing other types of bots. In this way, the researchers were able to cope, at least temporally, with the paradigm shift of bots\cite{cresci2017paradigm} and the effort of bot developers to evade detection techniques\cite{cresci2020decade}. This improvement corresponded to the third version of Botometer, available through its API until the end of August 2020. This version, highly used through its API by users \cite{yang2019arming}, was included in several studies \cite{rauchfleisch2020false} and considered as a state-of-art method for bot detection \cite{yang2019arming}. We use this tool in part of our experiments. 
Then, \cite{sayyadiharikandeh2020detection} introduces  Botometer version 4.  This research proposes an Ensemble of Specialised Classifiers. This approach consists of generating specific models for bot-operated accounts with different behaviors and then combine them through an ensemble and a voting system. It aims to deal with performance decrease when the training data present accounts with different behaviors. This alternative avoids retraining the model with a vast amount of data, which would be costly. Another problem that supervised methods may have is the lack of labeled data. \cite{kudugunta2018deep} presents a way to deal with this possible lack of data. This research uses data generation to create training data to feed a model that combines tweets' metadata with its content through an LTSM neural network. Using language-related features may provoke performance reduction when the models evaluate accounts interacting on other languages. 
Models  in \cite{knauth2019language} and  \cite{lundberg2019towards} address this issue, focusing on building language-independent models. The model in \cite{knauth2019language} used the tweets ' metadata to determine if an account is a bot or human. The research in  \cite{lundberg2019towards}  also introduces one method that is language-independent, which uses expressive account-based and content-based features. Others setbacks that can face supervised models are interpretability and noisy training data. Interpretability is an issue in ML algorithms, which may fall in the black-box metaphor, not letting humans understand the intermediate processes between an input and an output. The study in \cite{loyola2019contrast} approaches this issue extracting the features applying the contrast-pattern technique on aspects of the accounts such as usage, information, content-sentiment, or tweet content. Through this method, the model implemented is interpretable, enabling humans to understand why an account is classified as bot or human. Data noise in training data is a problem that may provoke a reduction of performance in a bot detector. \cite{yang2020scalable} uses a data selection technique to tackle this. This technique consists of choosing a subset of training data to optimize the performance of the model. It is an excellent method to maximize the existing available resources giving optimal results. Besides, in this research, it is presented a scalable classifier with 20 features. Scalability is essential when analyzing OSN because of the high volume of data. For our experiments, we make use of this method. Research in \cite{nasim2018real} also introduces a scalable supervised model.  It utilizes partial information of an account and its corresponding tweet history to detect content polluters in real-time.

As previously mentioned, bot detection is an evolving field because as soon as a new method appears, malicious bot developers work to beat it. Intending to detect the evolving trend of bots exposed in \cite{cresci2017paradigm}, research in \cite{cresci2017social} introduces the Social Fingerprinting technique. Social Fingerprinting models the online behavior of an account using the Digital DNA. Digital DNA is a string that encodes the different types of account interactions. Research in  \cite{cresci2017social} presents how to exploit Social Fingerprinting in both a supervised and unsupervised fashion using Lowest Common Substring(LCS) as a similarity measure between DNA strings. \cite{cresci2019cashtag} utilizes the former method to overcome a bot detection analysis over stock microblogs on Twitter.    \cite{kosmajac2019twitter} and \cite{pasricha2019detecting} present supervised models that uses Digital DNA.  \cite{kosmajac2019twitter} employs Statistical Measures for Text Richness and Diversity to extract the features from the Digital DNA. \cite{pasricha2019detecting} applies a lossless compression algorithm to the DNA string to obtain compression statistics as features. These features allow separating bot accounts and human-operated accounts, even permitting to visualize the division. 
Part of our work aims to combine this method with the data selection technique to build a robust method to detect bots across several domains.

Existing literature studied bot presence during the Covid-19 pandemic, such as \cite{ferrara2020types}. The study described and compared the behavior and discussion topics of bots and humans. Alternatively, other works analyzed the discourse during the Covid-19 pandemic on Online Social Networks(OSN). For instance, \cite{kleinberg2020measuring} and \cite{gao2020mental} studied emotional and sentiment dynamics on social media conversation around pandemic related topics.

\section{Implementation of bot detection models and comparison}

In this section, all the details about bot detection are explained. First, it is exposed how the features for bot detection were obtained and the different sets of features used. Then, the datasets used for training and test are presented. Moreover, the accounts from all the datasets are represented regarding a set of the features computed for bot detection. Finally, a comparison is made between the results of the different models implemented using a data selection technique and Botometer.

\subsection{Feature engineering}
\label{section:fengineering}

The features that we use for bot detection model can be split into two groups:  those obtained and derived from the metadata of each account and the variables obtained through the Social Fingerprint technique using compression statistics. 

The first approach consists of using as features for detection the metadata of each account,  and new variables derived from the raw metadata. The metadata is retrieved from the User Object related to each account. The features retrieved directly from the User Object are:

\begin{itemize}
\item \underline{\emph{statuses\_count}}: number of  tweets posted, including retweets.
\item \underline{\emph{followers\_count}}: number of followers.
\item \underline{\emph{friends\_count}}: number of accounts followed.
\item \underline{\emph{favourites\_count}}:  number of tweets liked by the account. 
\item \underline{\emph{listed\_count}}: number of public lists in which the account is involved. 
\item \underline{\emph{default\_profile}}: boolean indicating if the profile's theme or background has been altered.
\item \underline{\emph{verified}}: boolean indicating that the user has a verified account.
\end{itemize}

To compute some derived features from the metadata, the variable \emph{user\_age} is used. \emph{user\_age} corresponds to the difference in hours between the creation time of the last tweet accessible (probe time) and the creation time of the user\cite{yang2020scalable}. The features derived from the metadata of the User Object are:
\begin{itemize}
    \item \underline{\emph{screen\_name\_length}}: length of screen name string.
    \item \underline{\emph{num\_digits\_in\_screen\_name}}: number of digits in screen name string.
    \item \underline{\emph{name\_length}}: length of name string. 
    \item \underline{\emph{num\_digits\_in\_name}}: number of digits in name.
    \item \underline{\emph{description\_length}}: length of description string.
    \item \underline{\emph{friend\_growth\_rate}}: friends\_count/user\_age
    \item \underline{\emph{listed\_growth\_rate}}: listed\_count/user\_age
    \item \underline{\emph{favourites\_growth\_rate}}: favourites\_count/user\_age
    \item \underline{\emph{tweet\_freq}}: statuses\_count/user\_age
    \item \underline{\emph{followers\_growth\_rate}}: followers\_count/user\_age
    \item \underline{\emph{followers\_friend\_ratio}}: followers\_count/friend\_count
    \item \underline{\emph{screen\_name\_likelihood}}: It corresponds to the geometric mean of the likelihood of all bigrams in a screen name. More than 2 million unique screen names from random accounts of Twitter were retrieved to compute the likelihood of each one of the 3969 bigrams which can be created using the characters allowed in the screen name (Upper and lower cases letters, digits and underscore).
\end{itemize}

The intuition behind \emph{screen\_name\_likelihood} is that the screen name of bot operated accounts sometimes are constituted by a random string \cite{yang2020scalable}, being this a distinctive characteristic from humans.

The second approach, Social Fingerprinting, is a technique that consists of modeling the behaviour of an account through the Digital DNA, which is a string of characters based on the sequence of actions of a Twitter account. This string is produced encoding the behaviour through a mapping between the sort of interactions and characters or bases producing a DNA string. These bases form a set of unique characters called the alphabet. The alphabet is used to generate a sequence represented by a row vector or string which encodes a user behaviour \cite{cresci2017social}. More formally, an alphabet $\mathbb{B}$ is defined as \cite{pasricha2019detecting}
\begin{equation*}
\mathbb{B}=\{B_1,B_2,\ldots,B_N\}\quad B_i\neq B_j\quad \forall i,j=1,\ldots,N \wedge i\neq j 
\end{equation*}
which is utilised to generate a sequence whose expression is
\begin{equation*}
    s=\left(b_{1}, b_{2}, \ldots, b_{n}\right)=b_{1}b_{2}\ldots b_{n}\quad b_{i} \in \mathbb{B} \quad \forall i=1, \ldots, n
\end{equation*}

For our experiments, the following alphabet is used to encode a Twitter user behaviour:
\begin{equation*}
\mathbb{B}_{t y p e}^{3}=\left\{\begin{array}{l}
A \leftarrow \text { tweet } \\
\mathrm{C} \leftarrow \text { reply } \\
\mathrm{T} \leftarrow \text { retweet }
\end{array}\right\}=\{\mathrm{A}, \mathrm{C}, \mathrm{T}\}
\end{equation*}

The behaviour of a Twitter account is captured through its timeline and it is utilised to generate the DNA sequence. For instance, if an account $x$  first did a retweet, then two tweets and finally a retweet, its sequence utilising $\mathbb{B}_{t y p e}^{3}$ is $TAAT$. From here, it is implied that the length of the sequence depends on the number of tweets which are considered. In our case, we retrieved the maximum possible number of tweets(including retweets and replies) for each account, having the 3200 most recent tweets as a limit because of Twitter API restrictions\cite{TwitterA32:online}. The accounts which are protected or not possess any timeline cannot be analysed with this methodology.

The DNA sequences generated from the timelines are compressed using a lossless compression algorithm. Then, we compute the following features  \emph{original size of DNA string}, \emph{compressed size of DNA string} and \emph{compression ratio} ($\text{\emph{original DNA size}}/\text{\emph{compressed DNA size}}$). 

For our experiments we use the set of features listed below:
\begin{itemize}

    \item The features extracted and derived from the User Object previously introduced. This set of features is denoted as \emph{Light}.
    \item The original size of the DNA string and the compressed size of the DNA string. This set of features is referred as \emph{A}.
    \item The original size of the DNA string and the compression ratio. This set is denoted as \emph{B}.
    \item The compressed size of the DNA string and the compression ratio. This set is referred as \emph{C}.
    \item The original size of the DNA string, the compressed size of the DNA string and the compression ratio. This set is denoted as \emph{D}.
\end{itemize}

The set \emph{light} corresponds to the features used for bot detection in \cite{yang2020scalable} with the exception that it is not included the feature \emph{profile\_use\_background\_image} since it has been deprecated from the Twitter API \cite{Userobje99:online}. This set of features  allows implementing a scalable bot detection technique since each tweet retrieved with the Twitter API (versions 1.1 and Gnip 2.0)\cite{twitdev} contains the User Object of the corresponding account, with no need of obtaining extra data.\cite{yang2020scalable} However, this sort of approach can be vulnerable to adversarial attacks \cite{cresci2020decade}.  The set of features \emph{A}, \emph{B}, \emph{C} and \emph{D} are based on the research in \cite{pasricha2019detecting}. This technique provides a detection model which is more resistant against adversarial attacks \cite{pasricha2019detecting}, but scales worse. 

\subsection{Datasets}

In this section,  the datasets used for the implementation of the bot detection model are presented. Following the procedure in \cite{yang2020scalable}, we used some datasets for train and other datasets are set aside for testing. In this way, we expect to build a bot detection model that not just performs properly in cross-validation on the data used for training, but also generalises well when it is used for accounts displaying new behaviours, obtaining cross-domain validation. Most of the datasets have been obtained from  \url{https://botometer.iuni.iu.edu/bot-repository} or in other public repositories online.

The datasets used for training are:

\begin{itemize}
    \item \underline{\emph{Caverlee}}: To form this dataset, honeypots accounts were used to attract bot-operated accounts, mainly consisting of spammers, malicious, promoters, and friend infiltrators. This dataset was presented in \cite{lee2011seven}. 
    \item \underline{\emph{Cresci-17}}: The dataset was constructed using human annotators, labeled accounts from other datasets, and bot accounts purchased in online markets. The bots in this dataset include retweets spammers for political campaigns, hashtags spammers, URL spammers, job promoting bots, fake followers, and URL scammers. The dataset is used in \cite{cresci2017paradigm}.
    
    \item \underline{\emph{Varol}}:  The dataset was built annotating several accounts manually from different deciles of Botometer scores. It was first used in \cite{varol2017online}.

    \item \underline{\emph{Pronbots}}: The dataset was first shared in GitHub by a researcher in May 2018. The bots are Twitter advertising scam sites. It was used in \cite{yang2019arming}.
    
    \item \underline{\emph{Political}}: It consists of politics-oriented bots that were shared by Twitter user @john\_emerson. It was extracted from \cite{yang2019arming}.
    
    \item \underline{\emph{Botometer-feedback}}: It is made of those accounts which were annotated manually after been reported by Botometer users.  It is used in \cite{yang2019arming}.
    
    \item \underline{\emph{Vendor-purchased:}} It is uniquely composed of bots that play the role of fake followers. These accounts were bought by researchers from several companies. This dataset is used in \cite{yang2019arming}.
    \item \underline{\emph{Celebrity}}: This dataset, composed uniquely by human accounts, was extracted from \cite{yang2020scalable}. It was created by selecting Twitter accounts from celebrities.
    
\end{itemize}

\begin{table}[ht!]
\centering
\begin{tabular}{|l|l|l|l|l|}
\hline
\multirow{3}{10em}{\textbf{Training datasets}} & \multicolumn{2}{l|}{\textbf{User Object}} & \multicolumn{2}{l|}{\begin{tabular}[c]{@{}l@{}}\textbf{Social} \\ \textbf{Fingerprinting}
\end{tabular}} \\ \cline{2-5} 
                         & \emph{Human}           & \emph{Bot}            & \emph{Human}                                     & \emph{Bot}                                       \\ \hline
botometer\_feed          & 347             & 108            & 337                                       & 107                                       \\ \hline
varol                    & 1525            & 690            & 1331                                      & 668                                       \\ \hline
political                & 0               & 13             & 0                                         & 13                                        \\ \hline
cresci\_17               & 2907            & 5925           & 2440                                      & 5607                                      \\ \hline
celebrity                & 5814            & 0              & 5763                                      & 0                                         \\ \hline
vendor                   & 0               & 731            & 0                                         & 718                                       \\ \hline
pronbots                 & 0               & 1899           & 0                                         & 1723                                      \\ \hline
caverlee                 & 15211           & 14619          & 12824                                     & 14156                                     \\ \hline
\end{tabular}
\caption{Number of bot and human accounts in each training dataset. It is displayed the number of accounts for the cases where we use the features from the User Object and the Social Fingerprint.}
\label{table:train_datasets}
\end{table}

The datasets used for test are:

\begin{itemize}
    \item \underline{\emph{Botwiki}}: This dataset consists of 704 bot operated accounts. It is formed from active Twitter bots from \url{botwiki.org}. On this website, internet users can find an archive with self-identified bots. It is utilised in the research conducted in \cite{yang2020scalable}.
    \item \underline{\emph{verified}}: It is composed of human-verified user accounts extracted through the Twitter streaming API. It is utilised in \cite{yang2020scalable}.
    \item \underline{\emph{Rtbust}}: The dataset was created manually annotating the retweets retrieved from the last 12 days of June 2018. It was extracted from \cite{mazza2019rtbust}.
    \item \underline{\emph{Stock}}: The bot operated accounts were detected through similarities in timelines of accounts that contain tweets with specific cashtags in a five months period in 2017. In \cite{cresci2018fake} and \cite{cresci2019cashtag}, it is found the study through which the bot-operated accounts were detected and details about these accounts. The bots in this dataset present a coordinated behaviour.
    \item \underline{\emph{Gilani}}: The dataset was formed retrieving accounts with the Twitter Streaming API and splitting them into four groups regarding its followers. Then,  accounts from each group were extracted and annotated manually. The dataset was used in \cite{gilani2017bots}.
    \item \underline{\emph{Midterm}}: The dataset is composed of accounts that interacted during 2018 U.S. Midterm elections. The accounts were manually annotated as bot and human through the correlation between the tweeting timestamp and creation timestamp. The dataset is utilised in the research conducted in \cite{yang2020scalable}.    
    \item \underline{\emph{Kaiser}}: The accounts labeled as human correspond to those belonging to American and German politicians under the assumption that all are human-operated. On the other hand, the bot operated accounts are manually annotated for German accounts and extracted from \url{botwiki.org} in the case of English bots. This dataset was used in \cite{rauchfleisch2020false}.
\end{itemize}

The \emph{botwiki} and \emph{verified} datasets are considered together during the test as the \emph{botwiki-verified}. It is worth to mention that the datasets used for training are the same that in \cite{yang2020scalable}, whilst for testing, the datasets \emph{stock} and \emph{kaiser} are added to the datasets already used in \cite{yang2020scalable}. Including two more datasets for testing, we want to test the models with other bots with different natures.

In Table \ref{table:train_datasets} and Table \ref{table:test_datasets} the number of bot and human accounts which constitutes each dataset for the train and test is displayed . The tables are divided between user object and Social Fingerprinting because, as mentioned before, it is not possible to use DNA methods with those accounts which are protected or do not have timeline. Even though there are differences in the number of accounts in most of the datasets, these differences are thought not to be big enough to be misleading when the user object and Social Fingerprint approaches are compared.

\begin{table}[ht!]
\centering
\begin{tabular}{|l|l|l|l|l|}
\hline
\multirow{3}{10em}{\textbf{Test Dataset}} & \multicolumn{2}{l|}{\textbf{User Object}} & \multicolumn{2}{l|}{\begin{tabular}[c]{@{}l@{}}\textbf{Social} \\ \textbf{Fingerprinting}\end{tabular}} \\ \cline{2-5} 
                              & \emph{Human}           & \emph{Bot}            & \emph{Human}                                      & \emph{Bot}                                      \\ \hline
Rtbust                        & 332             & 321            & 317                                        & 314                                      \\ \hline
Gilani                        & 1418            & 1043           & 1293                                       & 997                                      \\ \hline
Kaiser                        & 1007            & 290            & 959                                        & 232                                      \\ \hline
Botwiki-verified              & 1985            & 685            & 1974                                       & 610                                      \\ \hline
Midterm                       & 7416            & 37             & 7290                                       & 32                                       \\ \hline
Stock                         & 6132            & 6964           & 5333                                       & 6246                                     \\ \hline
\end{tabular}
\caption{Number of bot and human accounts in each training dataset. It is displayed the number of accounts for the cases where we use the features from the User Object and the Social Fingerprinting.}
\label{table:test_datasets}
\end{table}

\subsubsection{Visualizing the datasets through compression statistics}

\begin{figure}[ht!]
\centering
\includegraphics[width=0.65\textwidth]{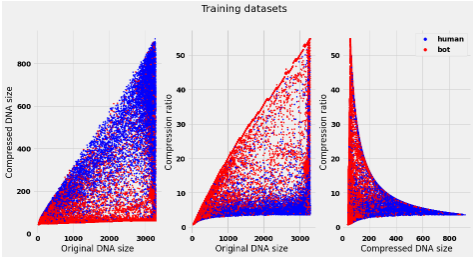}
\caption{Scatterplot representing accounts in train datasets through compression statistics.}
\label{fig:training_comprestion}
\end{figure}

\begin{figure}[ht!]
\centering
\includegraphics[width=0.8\textwidth]{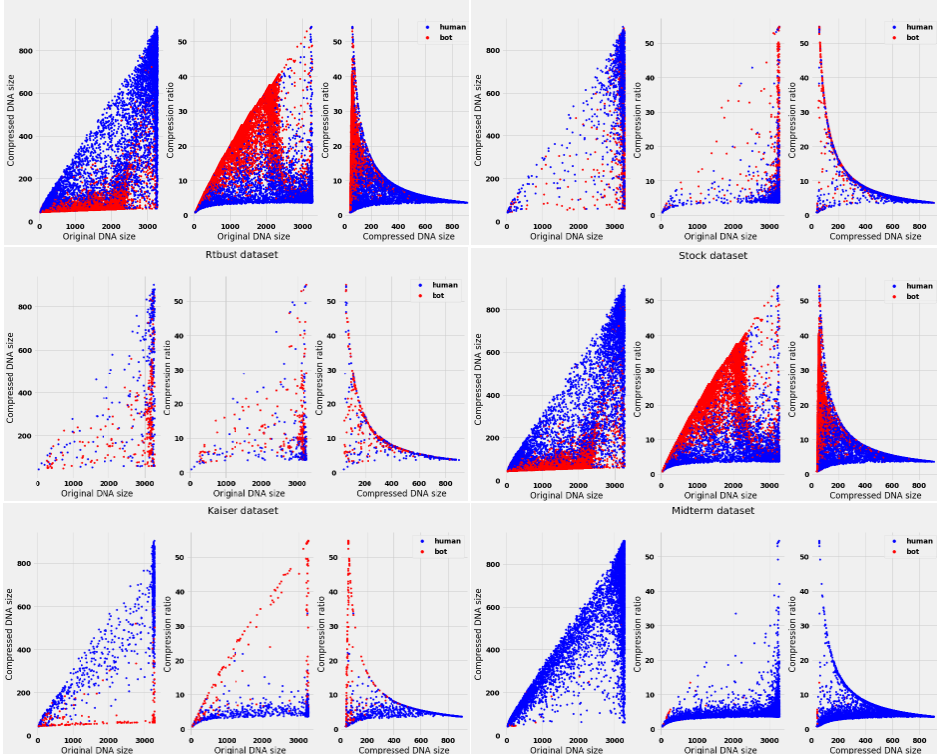}
\caption{Scatterplot representing accounts in test datasets through compression statistics.}
\label{fig:test_compression}
\end{figure}

Following the approach of  \cite{pasricha2019detecting}, we elaborate 2-D scatterplots representing the accounts in the datasets used in our work through the compression statistics. Figure \ref{fig:training_comprestion} displays all the datasets used for training represented by the three combinations of compression statistics. Figure \ref{fig:test_compression} conveys the same with each one of the test datasets. 

These plots intend to show that these features are not just useful to separate humans from the bots from a specific dataset, but it can be generalised to more cases. In fact, in most of the datasets, it is observed that there is a division between the bot and human-operated accounts.  

Besides, it is worth to mention the case of \emph{stock} dataset. In this dataset, the bots have a coordinated nature that makes inconvenient feature-based classifiers to detect them\cite{yang2020scalable}. However, looking at the scatterplot it seems that compression statistics achieve to separate both types of accounts. These plots can give us hints about the predictive power of models using these features for detection.

\subsection{Comparison between different Bot detection models using Data Selection}
\label{sec:comparison}

According to learning theory, using as much data as possible to train a model will provide the best models if the following conditions are met \cite{yang2020scalable}:

\begin{itemize}

\item The labels of the train data are correct.
\item The data considered is independent and identically distributed in the feature space.
\end{itemize}

In case these conditions are not met,  a  data selection method can be employed.  This method aims to encounter a subset of training data that will optimise the cross-validation performance on the training data and the ability of generalization on unseen data\cite{yang2020scalable}.   Data selection techniques have shown satisfactory results in different domains with data with noise and contradictory labels\cite{wu2007data} \cite{erdem2010ransac} \cite{zhang2014agreement}. The data selection technique will be used over the training data. Specifically, all the different combinations of train datasets are used, which supposes 247 different combinations.


Then, for each combination of datasets, each one of the sets in section \ref{section:fengineering} is used with the following classification algorithms: \emph{Logistic Regression}, \emph{AdaBoost}, \emph{Support Vector Machine with Linear Kernel}, \emph{Random Forest}, \emph{Gradient Boosting}, \emph{K Nearest Neighbors (KNN)}, \emph{Naive Bayes}, \emph{Multilayer Perceptron (MLP)}. Each possible combination is evaluated in all the test datasets using the AUC score. Using several classification algorithms, we intend to make a more intensive search to find the best performing model in \cite{yang2020scalable}, not just using combinations of datasets but also adding classification algorithms to the equation.

The MLP is composed by one hidden layer in the case of the set of features \emph{A},\emph{B},\emph{C} (120 neurons) and \emph{D}(150 neurons), and two hidden layers in the case of \emph{Light}(300 and 200 neurons). We use the default hyperparameters of the library sklearn for the other algorithms.

For the rest of the section, we will denote model as the vector of the form 
\begin{equation*}
    (x,y,z);\quad x\in X , y\in Y , z\in Z
\end{equation*}
where $X$ corresponds to the set composed by the 247 possible combinations of training datasets, $Y$ is the set formed by all the classification algorithms and $Z$ is the set formed by the set of features $Light$, $A$, $B$, $C$, $D$. 

We created 9880 different models, based on 247 train datasets, 8 algorithms, and 5 sets of features. Through our heuristic process, we selected 5 of them, i.e. the best model for each set of features. The process is the following:

\begin{enumerate}

\item  We group the models by feature set (obtaining 5 groups), and in each group we validate each of the 247$\times$8=1976 models against all the 6 test sets (AUC score).
\item We create a ranking for each test set in each group (6 rankings per group; in each one every model gets a value in the range 1..1976 based on its AUC score), and then we compute the sum of the 6 ranking values obtained by each model (1976 sums per group)
\item For each algorithm in each group, we take the model which has the lowest sum of the rankings (8 models per group)
\item  In each group we select manually the best performing model out of the 8 lowest-sum-models based on the AUC scores on the test sets and on its 5-fold cross-validation value. We based our selection primarily on the results of the test, always looking that the model performs well overall. However, in the case of similar results on the test datasets and considerable difference in the cross-validation (around 8\%) or slight signs of overfitting, we prioritize the cross-validation.

\end{enumerate}

This heuristic will provide a model that is not the best in every single test but works properly in all the test datasets. In this way, stability in applications is ensured.\cite{yang2020scalable}
 
 \begin{table}[ht]
\centering
\begin{adjustbox}{width=1.2\textwidth,center=\textwidth}
\small
\begin{tabular}{|l|l|l|l|l|l|l|l|l|l|}
\hline
\multirow{3}{3em}{Feature set}&\multirow{3}{5em}{Model}&\multirow{3}{5em}{Training dataset}&\multicolumn{7}{c|}{AUC Scores} \\
\cline{4-10}

                                                                                &                                                                &                                                                                             & \textit{Rtbust} & \textit{Gilani} & \textit{Kaiser} & \textit{\begin{tabular}[c]{@{}l@{}}Botwiki-\\ verified\end{tabular}} & \textit{Midterm} & \textit{Stock} & \textit{5-fold} \\ \hline
Light                                                                            & \begin{tabular}[c]{@{}l@{}}Gradient\\ Boosting\end{tabular}    & \begin{tabular}[c]{@{}l@{}}botometer\_feed,\\ varol,\\ cresci\_17,\\ celebrity\end{tabular} & 0.613           & 0.631           & \textbf{0.944}           & \textbf{0.991}                                                                & \textbf{0.964}            & 0.631          & 0.961           \\ \hline
A                                                                               & \begin{tabular}[c]{@{}l@{}}Logistic\\ Regression\end{tabular}  & \begin{tabular}[c]{@{}l@{}}political,\\ cresci\_17\end{tabular}                             & 0.700           & 0.719           & 0.943           & \textbf{0.991}                                                                & 0.962            & 0.922          & 0.957           \\ \hline
B                                                                               & \begin{tabular}[c]{@{}l@{}}Gradient\\ Boosting\end{tabular}    & \begin{tabular}[c]{@{}l@{}}botometer\_feed,\\ political,\\ cresci\_17\end{tabular}          & \textbf{0.720}           & \textbf{0.726}           & 0.938           & \textbf{0.991}                                                                & 0.961            & 0.858          & \textbf{0.968}           \\ \hline
C                                                                               & \begin{tabular}[c]{@{}l@{}}Random\\ Forest\end{tabular}        & \begin{tabular}[c]{@{}l@{}}political,\\ cresci\_17\end{tabular}                             & 0.660           & 0.691           & 0.927           & 0.980                                                                & 0.944            & 0.863          & \textbf{0.968}           \\ \hline
D                                                                               & \begin{tabular}[c]{@{}l@{}}Logistic \\ Regression\end{tabular} & \begin{tabular}[c]{@{}l@{}}botometer\_feed,\\ cresci\_17\end{tabular}                       & 0.699           & 0.719           & \textbf{0.944}           & \textbf{0.991}                                                                & 0.962            & \textbf{0.924}          & 0.945           \\ \hline
\end{tabular}
\end{adjustbox}
\caption{Best model for each set of features with their 5-fold cross validation and their performance in each test set.}
\label{table:results_all}
\end{table}

In Table \ref{table:results_all}, the best models according to our heuristic for each set of features are shown, along with the AUC score of the models in each test dataset and  5-fold cross-validation. We observe that the models with the features obtained through Social Fingerprint outperform or obtain similar results that the \emph{Light} model in all the cases. The \emph{stock} dataset is where the DNA models outperform more evidently the \emph{Light} model, with the model with the set of features \emph{D} obtaining the best result. This is because the bots in the  \emph{stock} dataset showed a coordinated behaviour that makes a feature-based model as \emph{Light} not convenient for their detection \cite{yang2020scalable}, while as evidence shows the Social Fingerprint together with compression statistics is an effective method to detect bots with a coordinated behaviour. Besides, we observe that the data selection technique is efficacious since none of the best models for each set of features used all the train datasets.

\subsubsection{Comparison with Botometer}

We made a performance comparison of the best models with the sets of features \emph{Light} and \emph{D} with Botometer. Botometer is an online social tool for bot detection. For the experiments, Botometer version 3 was used, which was available until the end of August 2020 through its API. Botometer version 3 has been used in several studies in literature and it has even been contemplated as the state-of-the-art tool for the detection of bots in Twitter \cite{sayyadiharikandeh2020detection}. It is a supervised model, specifically, it uses a Random Forest as a classification algorithm. Botometer v3 uses more than 1000 features from each account related to different fields such as the content of the tweets, its sentiment, the network of the account, or the user metadata \cite{varol2017online}. This model has been trained in the following datasets: \emph{caverlee}, \emph{varol},\emph{cresci-17},\emph{pronbots},\emph{vendor},\emph{botometer-feed},\emph{celebrity} and \emph{political} \cite{varol2017online}.

The three models present some significant differences. Both Botometer v3 and the model \emph{Light} use features extracted from the account, whereas the model with \emph{D} needs to construct the Digital DNA from the timeline of an account for prediction. Another difference is the number of features that use each model to classify an account. While Botometer v3 uses more than 1000 features, the model with \emph{Light} utilises 19 features and \emph{D} uses 3. However, the main difference between all the models comes with scalability: while the model with \emph{Light} allows to analyse accounts at the same pace that the tweets are retrieved, the other models need to cope with Twitter API rate limits since they need to retrieve the timeline of each account for classification, making them not scalable for the Twitter streaming.
In this experiment, apart from the AUC score, the following metrics are used to measure the performance of each model: F1, Accuracy, Recall, Precision, and Specificity. To compute the previous metrics is necessary to set a classification threshold. In the case of the Botometer v3, following research \cite{luceri2019red}, 0.3 is used as the threshold to separate humans from bots. That is to say, if the probability of an account to be a bot is greater than 0.3, then it is classified as a bot. This probability will also be referred as bot score. In the case of the model with the set of features \emph{D} and \emph{Light}, as done in \cite{yang2020scalable}, it is used as threshold the bot score that maximizes the F1 metric, maximizing precision and recall simultaneously. 

\begin{table}[ht!]
\begin{adjustbox}{width=1.2\textwidth,center=\textwidth}
\small
\begin{tabular}{|c|c|cccccc|}
\hline
\multirow{2}{4em}{Test Dataset}& \multirow{2}{5em}{\quad Model}&\multicolumn{6}{c|}{Evaluation metrics} \\
\cline{3-8}

                                                                                 &                                 & \textit{AUC} & \textit{F1} & \textit{Accuracy} & \textit{Recall} & \textit{Precision} & \textit{Specifity} \\ 
                                                                                 \hline
\multirow{3}{4em}{\textit{Botwiki-verified}}                                       & Light                            & 0.990        & 0.916       & \textbf{0.960}             & 0.857           & \textbf{0.985}              & \textbf{0.995}              \\ \cline{2-2}
                                                                                 & D                               & \textbf{0.991}        & \textbf{0.917}       & \textbf{0.960}             & \textbf{0.936}           & 0.899              & 0.968              \\ \cline{2-2}
                                                                                 & Botometer                       & 0.922        & 0.785       & 0.905             & 0.685           & 0.920              & 0.980              \\ \hline
\multirow{3}{4em}{\textit{Gilani}}                                                 & Light                            & 0.631        & 0.274       & 0.615             & 0.172           & 0.681              & \textbf{0.941}              \\ \cline{2-2}
                                                                                 & D                               & \textbf{0.718}        & \textbf{0.508}       & \textbf{0.670}             & \textbf{0.390}           & \textbf{0.726}              & 0.886              \\ \cline{2-2}
                                                                                 & Botometer                       & 0.689        & 0.456       & 0.644             & 0.341           & 0.687              & 0.880              \\ \hline
\multirow{3}{4em}{\textit{Kaiser}}                                                 & Light                            & \textbf{0.944}        & \textbf{0.817}       & \textbf{0.919}             & 0.807           & \textbf{0.827}              & \textbf{0.951}              \\ \cline{2-2}
                                                                                 & D                               & \textbf{0.944}        & 0.683       & 0.837             & \textbf{0.901}           & 0.550              & 0.822              \\ \cline{2-2}
                                                                                 & Botometer                       & 0.829        & 0.594       & 0.827             & 0.572           & 0.617              & 0.899              \\ \hline
\multirow{3}{4em}{\textit{Midterm}}                                                & Light                            & \textbf{0.964}        & \textbf{0.176}       & \textbf{0.964}             & 0.784           & \textbf{0.099}              & \textbf{0.964}              \\ \cline{2-2}
                                                                                 & D                               & 0.962        & 0.051       & 0.859             & 0.875           & 0.027              & 0.859              \\ \cline{2-2}
                                                                                 & Botometer                       & 0.958        & 0.101       & 0.912             & \textbf{0.905}           & 0.054              & 0.912              \\ \hline
\multirow{3}{4em}{\textit{Rtbust}}                                                 & Light                            & 0.613        & 0.217       & 0.536             & 0.131           & \textbf{0.636}              & \textbf{0.928}              \\ \cline{2-2}

& D                               & \textbf{0.699}        & 0.451       & 0.567             & 0.357           & 0.612              & 0.776              \\ \cline{2-2}
                                                                                 & Botometer                       & 0.625        & \textbf{0.473}       & \textbf{0.584}             & \textbf{0.377}           & \textbf{0.636}              & 0.788             \\ \hline
\multirow{3}{4em}{\textit{Stock}}                                                  & Light                            & 0.631        & 0.375       & 0.495             & 0.285           & 0.548              & \textbf{0.732}              \\ \cline{2-2}
                                                                                 & D                               & \textbf{0.924}        & \textbf{0.819}       & \textbf{0.771}             & \textbf{0.960}           & \textbf{0.714}              & 0.549              \\ \cline{2-2}
                                                                                 & Botometer                       & 0.756        & 0.780       & 0.719             & 0.927           & 0.673              & 0.480              \\ \hline
\end{tabular}
\end{adjustbox}
\caption{Comparison in performance of Botometer v3 and the best models with the set of features \emph{Light} and \emph{D}. }
\label{table:big_comparison}
\end{table}

In Table \ref{table:big_comparison},  the performance of the three models is displayed. We observe that the model with the set of features \emph{D} performs consistently well overall, outperforming or obtaining similar results to the other two models. It is worth to mention the good performance of the model with \emph{D} in the \emph{stock} dataset, where it performs the best. This gives evidence that the compression statistics extracted from the Digital DNA can detect bots that behave coordinately as happens in \emph{stock}. Moreover, by combining \emph{D} with data selection is possible to build a classifier that can generalise properly in different domains. Alternatively, the model with \emph{Light}, except for the \emph{stock} dataset, produces similar results that the other models, on some occasions outperforming them.  Besides, it shows the best specificity in all cases and it is scalable. As expected the model with \emph{Light} does not perform properly in \emph{stock} because of the coordinated behaviour of the accounts\cite{yang2020scalable}.  In contrast, Botometer seems to be more robust against the bots in \emph{stocks}, probably because its features cover more aspects apart from the user metadata. Results also confirm that is possible to obtain competitive performance using just a small set of features, as models with \emph{Light} and \emph{D}, rather than a bigger one as Botometer.

\section{Case study: High scale bot detection over the Covid-19 pandemic}

Many studies suggest how bots would manipulate public debate. This behaviour  would be particularly dangerous in the context of global health emergency. We then  posit  a main  research question: 

\textit{To what extent bots try to push disturbing action during the Covid-19 pandemic, in general and in relation  to specific topics?} 

More specifically, 

\textit{What is their prevalence and volume of posts activity compared to that of human accounts?} 

\textit{Does they exhibit any difference in the sentiment of the posts they share compared to ones shared by humans?}

To answer these questions, we study the bot presence on specific topics during periods of the first months of the pandemic. Then, after the bot detection analysis, we present the differences in the discourse between humans and bots, focusing on sentiment and hashtags. Through sentiment analysis we estimate the public opinion on a certain topics and would also track  COVID-19-related exposure to negative content in online social systems caused by bots activities. 

As regards procedure, we used hashtags to identify the tweets which were related to the same topic. We considered that two tweets belong to the same topic if they contain the same hashtags or a subvariant of them. For instance, the tweets with hashtags COVID19, covid, Covid19, CovidPandemic belongs to the topic COVID.

The tweets used for the experiments of this section were extracted from  public datasets in \cite{chen2020tracking}\cite{781wef4220}\cite{banda}  or Kaggle datasets. These datasets are composed of extracting tweets through the Twitter Streaming API. The tweets extracted contain specific hashtags or keywords with their variants related to COVID-19, or belong to specific accounts such as the World Health Organization (WHO). Even though most of the datasets contained tweets in several languages, they are mostly composed of English tweets since the hashtags or keywords used to extract the tweets refer to English terms. This fact implies that the tweets are mostly related to events in English-speaking countries such as U.S. or U.K. These datasets, due to Twitter regulations, contain the IDs of the tweets. Therefore, it was necessary to hydrate those IDs using the twarc library \cite{twarc} to obtain the full tweet object. We only consider English tweets for our experiments.

The topics and periods that we consider in our experiments are listed below:
\begin{itemize}
    \item Topic WUHAN on 25th and 26th January 2020. 
    \item Topic OUTBREAK on 25th and 26th January 2020.
    \item Topic COVID on 28th and 29th March 2020.
    \item Topic LOCKDOWN on 10th May 2020. 
    \item Topic TRUMP from 4th February to 21th February 2020.
\end{itemize}

As studies suggest that social media discourses mirror offline events dynamics, these topics and periods were studied as they were considered as prone for the presence of bots as they reflect some controversial issues in people’s conversations. 

WUHAN and OUTBREAK refer to the pandemic beginning where the virus had rapidly spread in China and received names such as "Wuhan virus" or "Wuhan coronavirus". In this context, authorities canceled large-scale events such as the Spring Festival, and there were traveling restrictions for more than 30 million people. These facts constituted an unprecedented event \cite{January286:online}. Moreover, 15 Chinese cities suffered partial or full lockdowns to attempt to limit the spread of the coronavirus \cite{12Getcau93:online}. 

The COVID topic on 28th and 29th March coincides with Trump considering quarantining New York \cite{Coronavi54:online} as there was a shortage of equipment for health workers and hospitals were overloaded \cite{March28c80:online}\cite{March29c46:online}.  Moreover, the milestone of 2000 deaths in the US was overcome in these days \cite{March28c80:online}.  

In the scope of LOCKDOWN on 10th May, there was a high criticism raised from the first steps out of the lockdown proposed by UK Prime Minister, Boris Johnson.\cite{BorisJoh77:online} 

Finally, the TRUMP case refers to the management of the start of the pandemic by President Trump, which was highly-criticized. In this period, there were problems with the COVID testing in the U.S.\cite{TheUnite45:online}, making it difficult to stop the spread of the virus. Besides, little attention was given to the coronavirus in the State of Union on 4th February, where President Trump spent less than 30 seconds referring to the COVID-19 situation\cite{Thelostm67:online}. Moreover, during this time, the US government had to manage the Diamond Princess cruise situation, where it was criticized the conditions around the Americans in the ship during the month of February\cite{Warrenca40:online}.

\begin{table}
\centering
\begin{tabular}{|l|l|l|}
\hline
Topic    & Accounts & Tweets \\ \hline
OUTBREAK & 64602    & 82030  \\ \hline
WUHAN    & 103916   & 163723 \\ \hline
COVID    & 312034   & 414097 \\ \hline
LOCKDOWN & 26813    & 31052  \\ \hline
TRUMP  & 10144    & 26865  \\ \hline
\end{tabular}
\caption{Number of accounts and tweets for each one of the cases studied.}
\label{table:n_accounts_tweets}
\end{table}

Table \ref{table:n_accounts_tweets} displays the number of unique tweets and accounts considered by each topic after hydrating the tweets. We use these tweets for our experiments.

\subsection{Bot analysis: Proportion of bots and distribution of bot score}

\begin{figure}[ht]
 \makebox[\textwidth][c]{\includegraphics[width=1\textwidth]{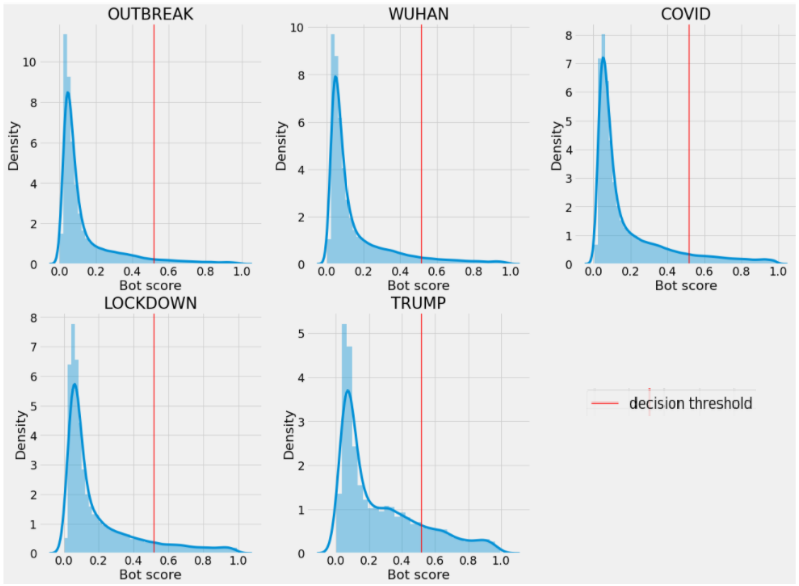}}
\caption{Bot score distribution for each of the cases studied.}
\label{fig:distribution}
\end{figure}

For the bot detection analysis, we use the model \emph{Light} as it displayed good results in section \ref{sec:comparison} and scalability. First, we study the distribution of the bot score in each one of the cases. The distributions are displayed in Figure \ref{fig:distribution}. The decision threshold corresponds to the one computed in \ref{sec:comparison}.  All the distributions are positively skewed, indicating a bigger presence of the human than bots. Moreover, except for the TRUMP distribution, it is observed a clear tail.

Then, we study if the distributions are similar between them. We run the Anderson-Darling statistical test to analyze if the samples of bot scores come from the same distribution. After running the test for all the pairs of distributions, we reject the null hypothesis at a 1\% significance level. We conclude that there is statistically significant evidence to state that the samples for each case do not come from the same distribution.

\begin{figure}[ht!]
\centering
\includegraphics[width=0.65\textwidth]{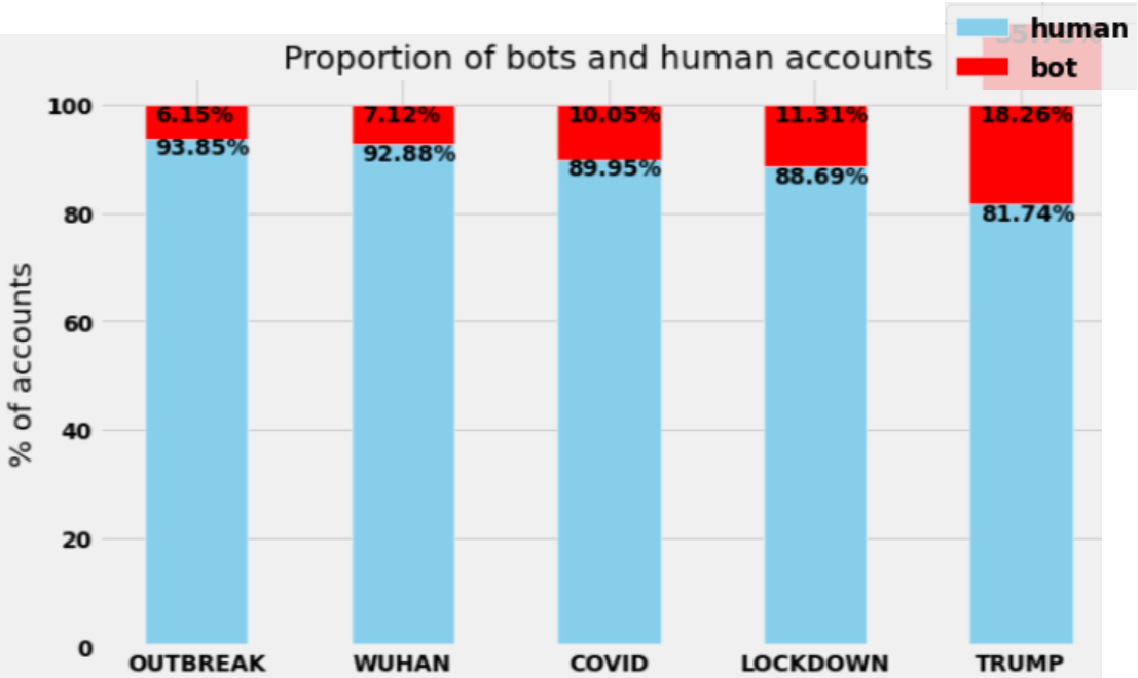}
\caption{Proportion of bot and human accounts that interacted in each case.}
\label{fig:proportion_accounts}
\end{figure}

Besides, we classify each account as a bot or human using the decision threshold computed in \ref{sec:comparison}. Figure \ref{fig:proportion_accounts} displays the proportion of bots and human accounts identified in each case. We notice that OUTBREAK and WUHAN cases have the smallest amount of bots, with only around 7\% bot-operated accounts. In COVID and LOCKDOWN, about 10\% and 12\% of the accounts are bots. The TRUMP case has the maximum proportion of bots with more than 18\%.

\begin{figure}[ht!]
\centering
\includegraphics[width=0.65\textwidth]{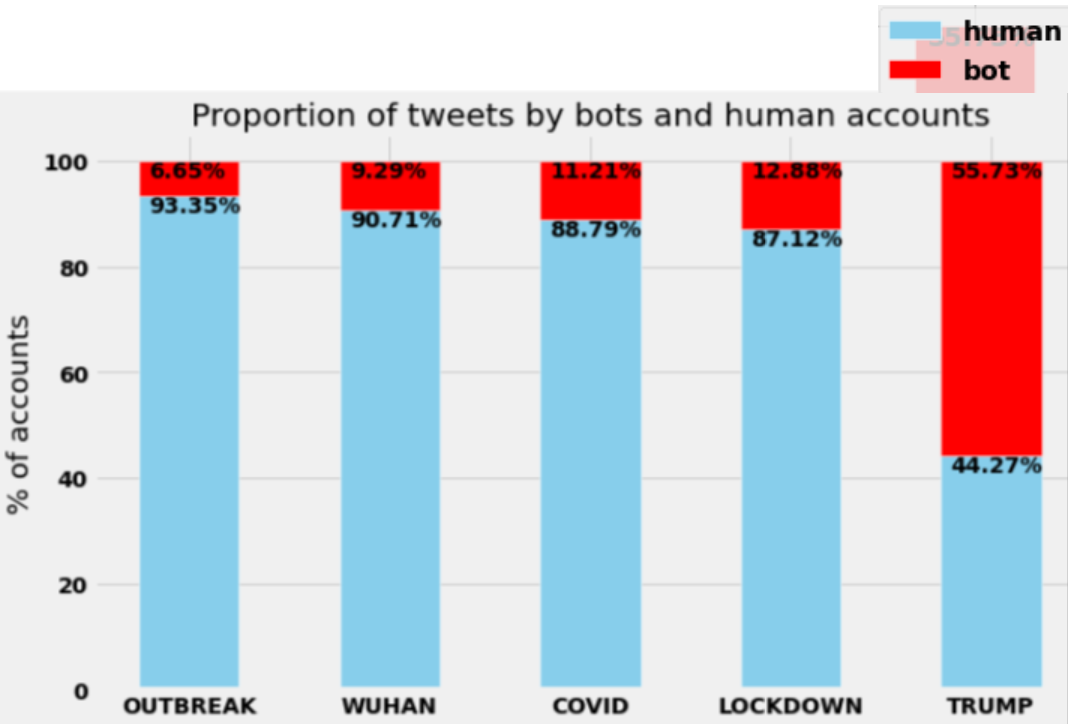}
\caption{Proportion of tweets which were produced by bots and humans in each of the cases studied.}
\label{fig:proportion_tweets}
\end{figure}

Then, we compute the number of tweets produced by bots and humans in each case. Figure \ref{fig:proportion_tweets} displays a comparative bar chart with the proportion of tweets created by bots and humans in each topic. We observe that in all the cases, except for TRUMP, the proportions of each type of account and tweets made by those accounts are analogous, not differing in more than 3\%. This fact indicates that bots and humans as a group present the same rate of activity in these cases. By contrast, in the TRUMP case, we see that bots are more active than humans. The bots, only 18.26\% of the accounts, produce the 55.73\% of total tweets in this case.


\subsection{Differences on the discourse: sentiment and hashtag analysis}

In order to understand whether bots would increase exposure to negative and inflammatory content in online social systems, we analyze tweets' content differences regarding bots and humans in each case. Sentiment analysis allows us to monitor social media to extract an overview of the opinion of Twitter users.

First, we implement sentiment analysis in each one of the situations using VADER. We analyze the sentiment to learn about the reactions of users in each one of the situations studied. Then, the sentiment analysis was extended for the LOCKDOWN and TRUMP cases, using only the hashtags in the tweets to predict tweets' sentiment. Eventually, we examine the most common hashtags for bots and humans and discuss differences between each group. 

\subsubsection{Sentiment Analysis using VADER}


We use VADER\cite{GitHubcj89:online} to implement the sentiment analysis for all the cases. VADER is a sentiment model specifically designed to analyze microblog-like contents as tweets. To predict the sentiment, VADER uses a list of lexical features with their corresponding gold-standard sentiment intensities, combined using a set of five grammatical rules. According to the study in \cite{ribeiro2016sentibench}, where it has been benchmarked more than 20 techniques using 18 datasets, VADER is one of the best sentiment analysis methods for Social Media messages. Apart from its performance, we choose VADER because of its scalability and its simple utilization. There is a VADER implementation available in the NLTK library\cite{bird2009natural}. Besides, it needs little preprocessing compared to other methods. We apply the following preprocessing steps to the tweet content before using the VADER sentiment analyzer:

\begin{enumerate}
    \item Remove extra white spaces.
    \item Remove links and/or URLs.
    \item Remove username.
    \item Remove RT symbol.
    \item Remove HTML elements.
    \item Remove \# symbol.
    \item Remove non-ASCII elements.
    
\end{enumerate}

We based our experiment on the output of VADER, denoted as compound score. This metric corresponds to a single unidimensional measure for the sentiment. It is the result of summing the score of each word in the lexicon, then adjusting this value regarding the grammatical rules and normalizing it. It ranges between -1, the most negative value, and 1, the most positive.

We use this compound score to label a tweet as positive, neutral or negative. Specifically, as recommended by VADER documentation \cite{gilbert2014vader}, we use the following 
thresholds:
\begin{itemize}
\item Positive: \emph{compound score $\geq$ 0.05}
\item Neutral: \emph{-0.05 $\leq$ compound score $\leq$ 0.05}
\item Negative: \emph{compound score $\leq$ -0.05}
\end{itemize}

\begin{figure}[ht!]
\centering
\makebox[\textwidth][c]{\includegraphics[width=1\textwidth]{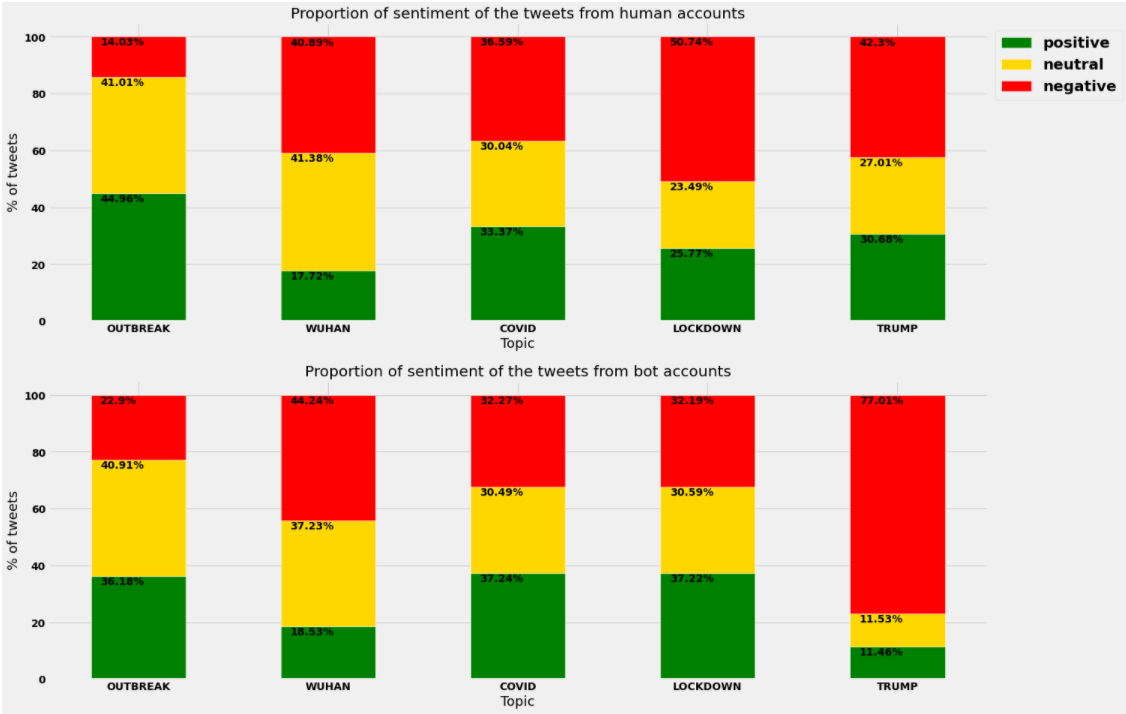}}
\caption{Sentiment of the tweets for each of the cases studied for human and bot operated accounts.}
\label{fig:normal_vader_proportion}
\end{figure}

Figure \ref{fig:normal_vader_proportion} displays the proportions of tweets for each case after using the sentiment thresholds above. 

We observe that the case OUTBREAK show similar proportions for bots and human. There is a greater presence of positive and neutral tweets (around 80\%), being the negative tweets the minority. 

Regarding WUHAN, we also notice similar proportions between humans and bots. In contrast to OUTBREAK, there is a bigger proportion of negative and neutral tweets, being the positive tweets the minority with only around 18\% for bots and humans. It is worth mentioning that even though WUHAN and OUTBREAK are highly related and it is considered the same period, they show inverse behaviors. 

Regarding COVID, we notice that both humans and bots produced similar proportions for negative, neutral, and positive tweets.  The former fact might indicate a division of users' opinion into the measure of quarantining New York.

Alternatively to the previous cases, we see that the humans and bots accounts show different proportions in the LOCKDOWN and TRUMP cases. 

In LOCKDOWN, bots show similar amounts of positive, neutral, and negative tweets. However, humans mainly display a negative tendency (50.74\% of the total tweets), while the positive and neutral correspond to half of the tweets in a balanced way. This value might indicate public opinion disagreement with the first steps out of the Lockdown proposed by the UK Prime minister. 

In the TRUMP case, we observe a more evident difference between the sentiment proportions of tweets produced by bots and humans. We notice that humans present a balance between the three classes with a little dominance of negative tweets (42\% negative-27\% neutral - 31\% positive). We interpret this result as a light dissent of users with President Trump's political performance during that period. On the other hand, negative-sentiment tweets correspond to the majority for bots, with almost 80\% of the tweets. These values represent a drastic difference, showing that tweets generated by bots have a predominantly negative attitude.

So far, we have used thresholds and discrete labels to measure the sentiment. However, one setback of this approach is the inability to count on intensities. For instance, we cannot differentiate between an extremely and slightly negative tweet since both are considered negative. To overcome this limitation and make a more extensive study, we complemented the previous analysis by studying the sentiment with a continuous metric, .i.e. the compound score. This analysis allows us to comment also about the intensity of the tweet content.  

\begin{figure}[ht!]
\centering
\makebox[\textwidth][c]{\includegraphics[width=1\textwidth]{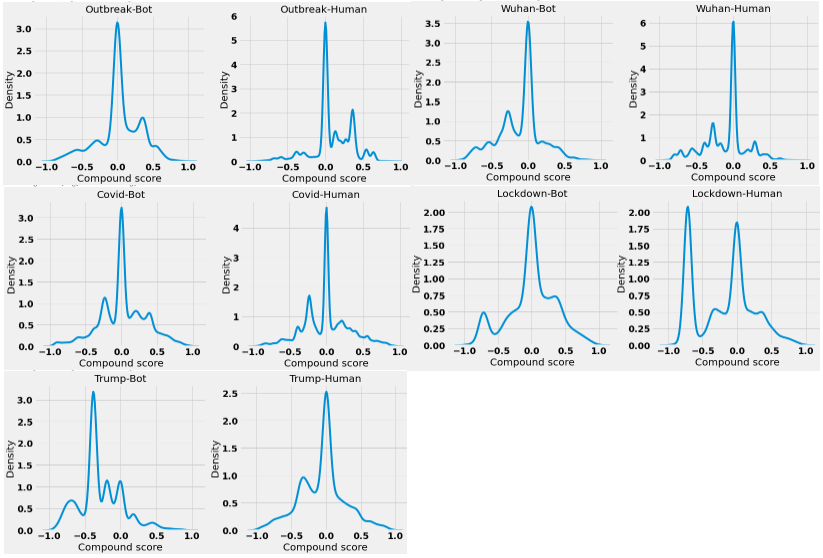}}
\caption{Distribution of sentiment compound score for each case regarding human and bot accounts.}
\label{fig:dist_compound_score}
\end{figure}

Figure \ref{fig:dist_compound_score} displays the distributions of compound scores regarding bots and human accounts for each case. We observe that for OUTBREAK, WUHAN, and COVID, the location of the peaks of the distributions for human and bots are similar. Moreover, most of the scores are around 0 in these cases, the samples not presenting extreme scores. In the human distribution in the LOCKDOWN case, we observe that the negative tweets display a more extreme score (peak between -0.6 and -0.8) than those positive (less than 0.5). This fact explains that human users were more drastic when they refer negatively to Lockdown than when they referred positively. Besides this case, it is the only distribution where we can notice two peaks, one in the neutral interval and one in the negative scores. Alternatively, regarding bots in the LOCKDOWN case, we observe that the positive tweets are close to the central scores, while we notice negative scores along the spectrum, from more neutral to more extreme scores. Concerning the TRUMP case,  bots distribution only displays a peak which shows that most tweets have a slightly negative sentiment. In the case of humans, all the compound scores are located in the center of the distribution. This fact implies that positive and negative tweets do not show extreme positions.

Furthermore, we run an Anderson-Darling test to see if the samples of the compound scores between humans and bots present the same distribution for each case. After running the test for all the pairs of distributions, we reject the null hypothesis at a 1\% significance level. Therefore, we conclude that there is statistically significant evidence to state that samples do not come from the same distribution.

The experiments in this subsection have some limitations. First, even though VADER presents the previously described advantages, it is not attuned for tweets that regard politics. This fact can reduce the performance of VADER on occasions. Besides, using hashtags to extract the tweets of the same topic might be sensitive to spam. Twitter users can use hashtags to gain popularity or attention, though it is not related to the tweet content. Moreover, our hashtag-based method for extraction can retrieve some tweets which are not fully-related to the topic we are studying. That being said, the limitations are not thought to be significant enough to not able to grasp valuable insights about the overall opinion displayed by the Twitter community about specific topics and analyze differences in sentiment between humans and bots.

\subsubsection{Sentiment Analysis using hashtags}

We evaluate the sentiment through the hashtags in the tweets. By doing so, we expect to overcome some of the limitations exposed in the previous section and make a more extensive analysis. Previously, manually labeling all the hashtags in the tweets as positive, negative, and neutral, we follow the following approach to obtain the sentiment of the tweets:
\begin{itemize}
    \item If a tweet contains at least one positive hashtag, the tweet is labeled as positive.
    \item If a tweet contains at least one negative hashtag, the tweet is labeled as negative.
    \item If a tweet contain does not contain positive nor negative hashtag, the tweet is labeled as neutral.
    \item If a tweet contain at least a positive hashtag and a negative hashtag, the tweet is labelled as inconclusive.
    
\end{itemize}

It is worth mentioning that all the tweets evaluated contain at least one hashtag because of the extraction method. Moreover, as results will convey, inconclusive tweets are a minority since a user will refer to negative or positive hashtags regarding a topic, not with both. 

In particular, we only evaluated the topics LOCKDOWN and TRUMP since they show a higher polarity. We expect to gain insights into the opinion of users regarding Trump's political performance and Lockdown measures. The hashtags were manually labeled following specific guidelines for each one of the cases.

We followed the rules below to label the hashtags in the LOCKDOWN tweets:

\begin{itemize}
\item It is assigned +1 (positive) to all hashtags which display a favourable attitude towards the lockdown and individual protection measures.
\item  It is assigned -1 (negative) to those hashtags against the lockdown and individual protection measures. 
\item The rest of the cases are labelled as 0 (neutral).
\end{itemize}
We followed the guidelines below to label the hashtags in the TRUMP tweets:

\begin{itemize} 
\item  It is assigned +1 (positive) to those hashtags in favour of Trump or its campaign, the GOP, or conspiracies theories who support the figure of Trump. Hashtags containing slogans pro-Trump are also labeled as 1.
\item It is assigned -1 (negative) to those hashtags which shows an offensive attitude towards Trump, including nicknames. It is also given -1 to those hashtags which were  against  GOP, constitutes sarcastic slogans, or are in favour of the democratic party. 
\item It is given 0 to the rest of the hashtags.
\end{itemize}

Using the previous instructions, in the LOCKDOWN case, we labeled 221 negative hashtags and 241 positive hashtags out of the 14376 in the LOCKDOWN tweets. Otherwise, in the TRUMP case, we obtained 938 negative hashtags and 367 positives out of 9678 total hashtags. Moreover, there were less than 1\% of inconclusive tweets for both cases.

\begin{figure}[ht!]
\centering
\makebox[\textwidth][c]{\includegraphics[width=0.65\textwidth]{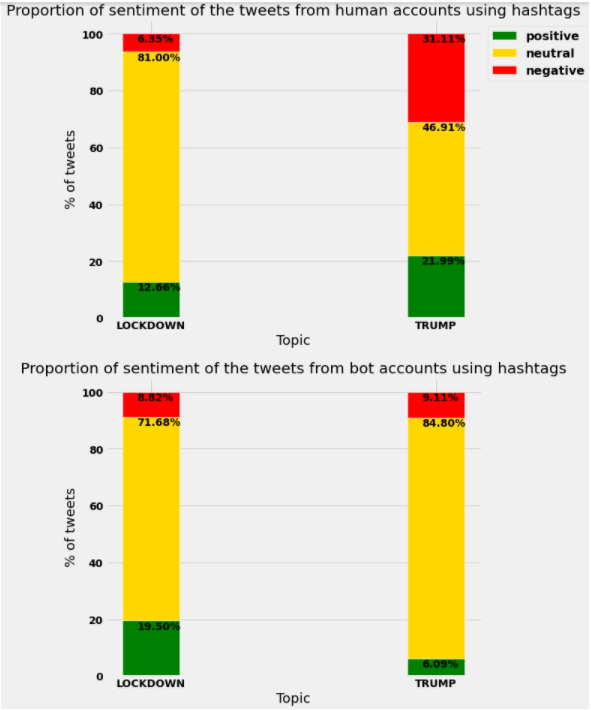}}
\caption{Sentiment of the tweets  using hashtags for human and bots for the LOCKDOWN and TRUMP case.}
\label{fig:hashtags_sentiment_plot}
\end{figure}

The results using the hashtag-based method are shown in Figure \ref{fig:hashtags_sentiment_plot}. We observe a predominant proportion of neutral tweets in all cases. This result matches with the nature of hashtags: they usually label tweets in a topic,  expressing an opinion being less frequent. However, when they express opinion they give us evidence of the position of the user. This fact allows us to gain more accurate insights into the opinion of the topics studied. In the LOCKDOWN case, we observe twice as many tweets with positive sentiment (12.66\%)  than tweets with negative sentiment (6.35\%). From these results, we could say that more people agree with the need for measures in favor of the lockdown than people who do not. We observe the same tendency regarding the bots in the LOCKDOWN case; it is bigger the proportion of positive tweets than negative. In both cases, the proportion of neutral tweets supposes the majority of tweets with 81\% for humans and 71.68\% for bots. For the TRUMP case, humans and bots display a bigger proportion of negative tweets than positive. However, the differences in proportions between one and the other differ significantly. For bots, the difference between positive and negative is 3\%, while neutral tweets constitute almost 85\% of the tweets. Concerning humans, we observe that less than 50\% of the tweets are neutral. We notice a bigger proportion of negative sentiment tweets than positive; 31\% against 22\%. This fact display that public opinion has a more negative attitude towards Donald Trump in that period.

\subsubsection{Hashtags analysis}

In this section, we explore the differences in the discourse regarding the hashtags in bots and humans. This analysis aims to see if bots and humans tweet about different things even in the same context. Significant differences in the hashtags between bots and humans would imply that conversations between humans and bots differ. To implement this analysis, we plot, for each case, the 20 most frequent hashtags used by humans and bots.

\begin{figure}[ht!]
\centering
\makebox[\textwidth][c]{\includegraphics[width=0.9\textwidth]{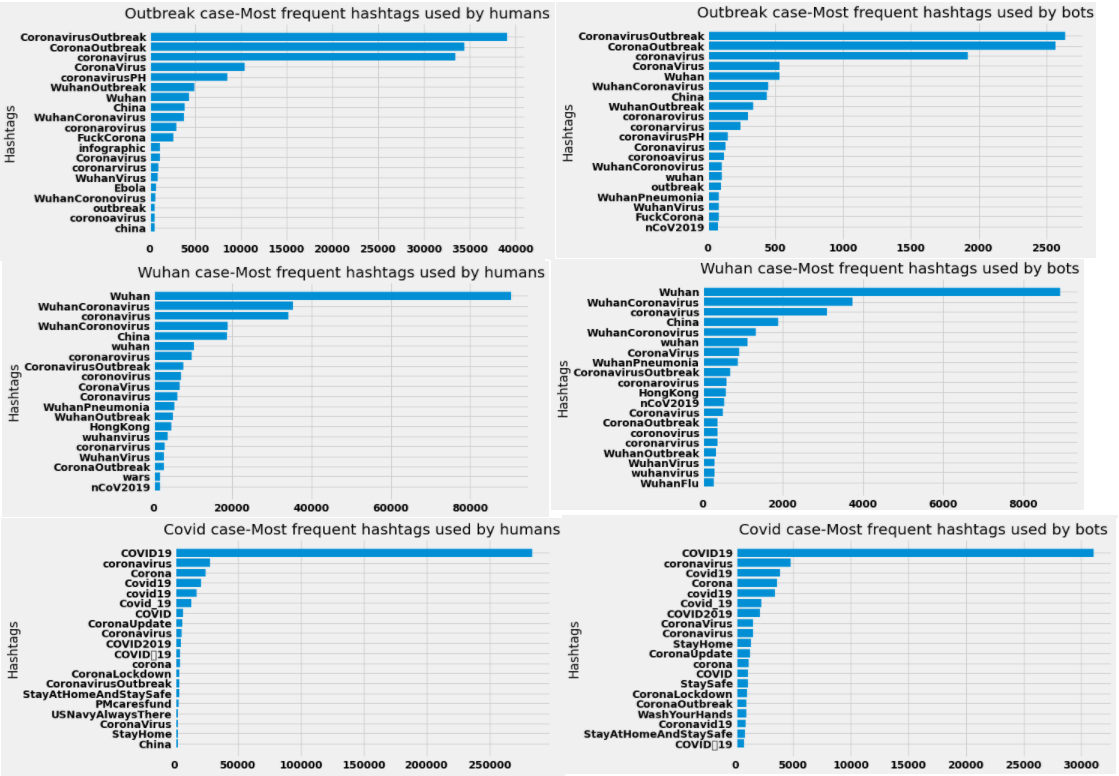}}
\caption{Most frequent hashtags for the OUTBREAK,WUHAN and COVID  cases.}
\label{fig:most_frequent_word_OWC}
\end{figure}

Figure \ref{fig:most_frequent_word_OWC} displays the most frequent hashtags that used humans and bots for the OUTBREAK, WUHAN, COVID cases. We observe in all three cases that humans and bots use similar hashtags, indicating very homogeneous discourse. We list below few differences that we can spot between the hashtags in each case:
\begin{itemize}
    \item In contrast to bots, \emph{\#infographic} or \emph{\#Ebola} are between the most common hashtags used by humans in OUTBREAK. The former might be because human users are sharing pieces of information based on infographics. The latter could mean that human users find similarities between the Ebola outbreak in Europe and U.S. in 2014 and the Covid-19 situation.
    \item In the WUHAN case, bots utilize the term \emph{\#WuhanFlu} to refer to COVID-19 in contrast to humans.
    \item In the COVID case, we can see support by human users to the U.S. Navy with the hashtags \emph{\#USNavyAlwaysThere}. This hashtag probably refer when the U.S. Navy sent a hospital ship to help the area of New York.\cite{Presiden41:online}. Conversely to bots, we observe that humans use \emph{\#PMcaresfund}. PM CARES Fund was created in India on 27th March to fight against Covid-19 and analogous pandemic situations in the future \cite{AboutPMC58:online}. On the other hand, bots in COVID share the message \emph{\#WashYourHands} as a prevention measure for Covid.  
\end{itemize}  

\begin{figure}[ht!]
\centering
\makebox[\textwidth][c]{\includegraphics[width=0.9\textwidth]{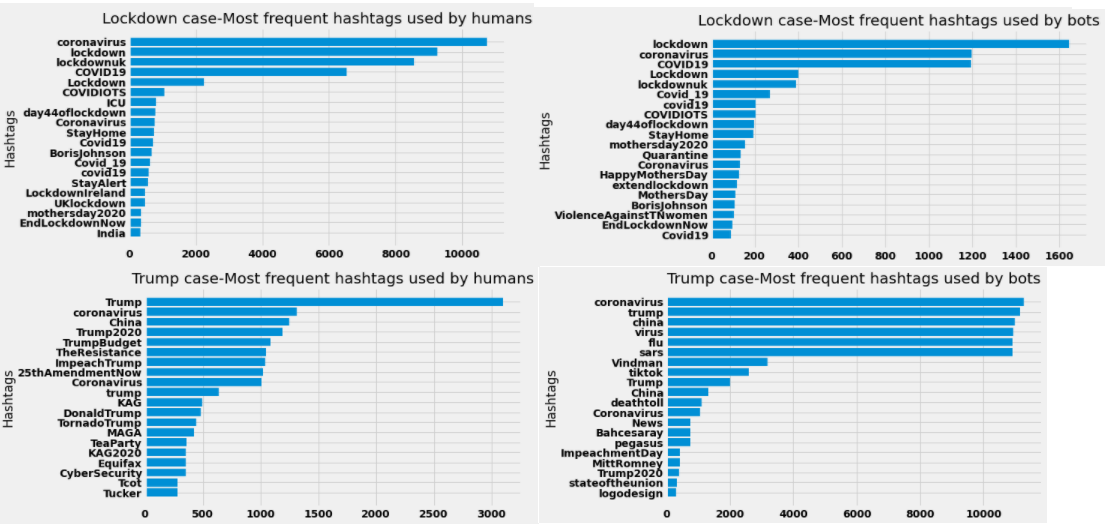}}
\caption{Most frequent hashtags for the LOCKDOWN and TRUMP cases.}
\label{fig:most_frequent_word_L_T}
\end{figure}

Figure \ref{fig:most_frequent_word_L_T} displays the most frequent hashtags that used humans and bots for the LOCKDOWN and TRUMP cases. We observe in LOCKDOWN that the most frequent hashtags are equal for bots and humans. In general terms, we can see hashtags referring to U.K., India, or South Africa events in both cases. For instance, \emph{\#lockdownuk} refers to the U.K. lockdown, and hashtags such as \emph{\#HappyMothersDay} are related to India. In India, Mother's Day is the second Sunday of May, which fell on 10th May in 2020 \cite{Mother:online}. Otherwise, \emph{\#day44oflockdown} regards South Africa, since the 10th May was the 44th day of lockdown in South Africa \cite{Coronavi79:online}. However, one difference between bots and humans in the discourse is that humans also focused on the Lockdown in Ireland as \emph{\#LockdownIreland}. Besides, humans use the hashtag \emph{\#ICU} on its discourse, probably referring to the pressure in U.K. hospitals for the high occupancy of the Intensive Care Unit in the U.K. \cite{2020051519:online}. In contrast to humans, we also notice that bots use the hashtag \emph{\#ViolenceAgainstTNwomen}, referring to the violence suffered by women in the Indian State of Tamil Nadu. 

The TRUMP case is where we observe a bigger difference between the discourse of humans and bots. One of the main differences we spot is the pro-Trump hashtag \emph{\#Trump2020}. We also notice some other pro-Trump hashtags such as \emph{\#KAG2020},\emph{\#KAG},\emph{\#MAGA}. Besides, the Tea Party movement (\emph{\#TeaParty}) and Top Conservatives on Twitter (\emph{\#Tcot}) should favor President Trump. It seems humans show more evidently their support to Trump than bot-operated accounts. One of the most recurring topics for humans is the budget proposal of Trump on 10th February. The proposal advocated for an increase in defense; and cuts and restrictions in foreign aid and social welfare programs\cite{Trumpsub60:online}. Humans refer directly to Impeachment with hashtags against Trump, such as  \emph{\#ImpeachTrump} and \emph{\#25thAmendmentNow}. Besides, humans mention the hacking attack on Equifax, which affected the data of 145 million Americans\cite{Equifaxb3:online}. On the other hand, we observe that bots use recurrently hashtags \emph{\#virus}, \emph{\#flu}, and \emph{\#sars} to refer to COVID-19 pandemic. Besides, we notice that bots also speak about the Impeachment, but they refer differently to it. They do not use hashtags that display opposition to Trump as humans. They utilize neutral hashtags as \emph{\#ImpeachmentDay}, or containing the name of people that participate in the process, as retired U.S. Army Lieutenant Colonel Alexander Vindman (\#Vindman) or Republican Senator Mitt Rodney (\#MittRodney). It is worth mentioning that both showed opposition to Trump during the Impeachment process \cite{WhyMittR84:online} \cite{Alexande71:online}. We also observe that bots referred to Tiktok platform (\emph{\#tiktok}) and the state of the union speech (\emph{\#stateoftheunion}). Moreover, we also perceive that some bots aim to spread news, such as the crash of a plane from Pegasus airline (\emph{\#Pegasus}) \cite{PegasusA25:online} or the avalanche in Bachcesaray, Turkey (\emph{\#Bachcesaray}) \cite{Dozensof86:online}.

To sum up, we observe that in the cases  OUTBREAK, WUHAN, and COVID exist few dissimilarities between the discourse of bots and humans regarding the hashtags analysis. However, these differences increase in the LOCKDOWN and TRUMP, being the latter case where humans and bots differ most in their discourse.

\section{Discussion and conclusions}

In this work, we produce a comparison between supervised Bot Detection methods using Data Selection and a case study related to the Covid-19 pandemic.

The comparative study aims to find a consistent model with the best balance between cross-validation and cross-domain generalization. In the comparison, we compared the method in \cite{yang2020scalable} with \cite{pasricha2019detecting}. We followed a similar pipeline to \cite{yang2020scalable}. However, we extended the study using an extra test dataset, the metadata currently available in Twitter API, and several classification algorithms. Besides, we applied the data selection technique to \cite{pasricha2019detecting}. The experiments displayed that combining the \cite{pasricha2019detecting} with data selection produce excellent results, not only outperforming the model from \cite{yang2020scalable} in certain situations but also compared to Botometer version 3.  The model implemented proves to be more effective than the other two when detecting bots that convey a coordinated behavior. Alternatively, the model with the approach from \cite{yang2020scalable}, after proving different classification algorithms, also produces competitive results. We use this model in the case study because of its performance and scalability.

In our case study, we set forth to investigate to what extent automated bots accounts were active on Twitter during the health global crisis due to Covid-19 pandemic. Prior works demonstrated how bots acted  massively in different context such as elections campaigns or Brexit crisis and how they have been used in malicious manners to spread misinformation and manipulating  public debate. This behaviour  would be particularly dangerous in the context of the  global health outbreak  when public  discourse goes more and more online due to social distancing measures.

Our findings paint a picture where while automated accounts are numerous and active when discussing some controversial issues, such as the lockdown measures in the UK or the pandemic beginning in WUHAN, usually they seem not increase exposure to negative and inflammatory content in online social systems. Despite this, when discourse switch to the management of the pandemic by President Trump, bots became more and more active in the spreading of discontent related to its  policy decisions as a consequence of the underestimation of the outbreak. In this case, sentiment-related values display a drastic difference, showing that tweets generated by bots have a predominantly negative attitude. 

By evaluating the sentiment  through the hashtag in the tweets, we expect to gain a deeper understanding into the opinion of bots and humans regarding Trump's political performance and lockdown measures. Concerning humans, we could say that more people agree with the need for measures in favor of the lockdown than people who do not. Consistently, Trump's policy of underestimating the health emergency has been heavily criticized by human users. However, in these cases we cannot definitely conclude that the bots are responsible for exposure to negative content related to these two topics. 

Furthermore, this result seems consistent with the hashtags analysis aims to explore the differences in the discourse regarding bots and humans. Significant differences in the hashtags shared by human and bots would imply that conversation between them differ.  While in the cases  OUTBREAK, WUHAN, and COVID exist few dissimilarities between the discourse of bots and humans, these differences increase in the LOCKDOWN and TRUMP cases, being the latter where humans and bots differ mostly in their discourse. Nevertheless in the TRUMP case it seems humans show more evidently their support to Trump than automated accounts, disproving the hypothesis, limited to the case study, of any conspiratorial attitude pushed by bots.

\bibliographystyle{plain}
\bibliography{main}
\end{document}